\documentclass[11pt]{article}

\usepackage[utf8]{inputenc}
\usepackage[T1]{fontenc}
\usepackage[margin=1in]{geometry}
\usepackage{graphicx}%
\usepackage{multirow}%
\usepackage{amsmath,amssymb,amsfonts}%
\usepackage{amsthm}%
\usepackage{mathrsfs}%
\usepackage{xcolor}%
\usepackage{textcomp}%
\usepackage{booktabs}%
\usepackage{authblk}%
\usepackage[round]{natbib}%
\usepackage{hyperref}%

\theoremstyle{plain}%

\providecommand{\keywords}[1]{\medskip\noindent\textbf{Keywords:} #1\par}

\begin{document}

\title{Injection-rate effects on failure in a fluid-saturated granular fault gouge}

\author[1]{Pritom Sarma\thanks{Corresponding author: \texttt{pritom.sarma@mail.huji.ac.il}}}
\author[2,3]{Stanislav Parez}
\author[4,5]{Renaud Toussaint}
\author[1,5]{Einat Aharonov}

\affil[1]{Institute of Earth Sciences, Hebrew University of Jerusalem, Jerusalem, Israel}
\affil[2]{Faculty of Science, Jan Evangelista Purkyn\v{e} University in \'Ust\'i nad Labem, \'Ust\'i nad Labem, Czech Republic}
\affil[3]{Faculty of Science, Charles University, Prague, Czech Republic}
\affil[4]{Universit\'e de Strasbourg, CNRS, Institut Terre et Environnement de Strasbourg, Strasbourg, France}
\affil[5]{PoreLab, The Njord Centre, Departments of Physics and Geosciences, University of Oslo, Oslo, Norway}

\date{}

\maketitle

\begin{abstract}
Fluid injection into the Earth's subsurface, performed for energy extraction, waste disposal, and resource development, is known to  reactivate gouge-filled faults and induce seismicity, a key hazard in modern geotechnical operations. Nevertheless, the role of injection rate in controlling fault-gouge failure remains poorly understood. Here we present both an analytical theory and coupled fluid--granular (discrete element) numerical simulations to explain this rate dependence. Assuming a pre-stressed gouge-filled fault subject to fluid injection, we derive a pore-pressure diffusion equation with a dilative sink. Its solution predicts a rate-dependent failure criterion, arising from pressure heterogeneity within the layer: slow injection allows pressure to diffuse uniformly throughout the layer, promoting uniform weakening, whereas rapid injection produces strong gradients, leaving distal regions stronger. The numerical simulations confirm the theory and reproduce experimental observations not captured by classical, uniform-pressure effective-stress theory. The framework links grain-scale physics to fault-scale failure and provides quantitative guidance for the design of injection protocols in geotechnical operations involving granular geomaterials.
\end{abstract}

\keywords{Granular geomaterials, Fault gouge, Fluid injection, Hydro-mechanical coupling, Discrete element method, Pore-pressure diffusion, Induced seismicity}

\section*{Highlights}
\begin{itemize}
\item A coupled granular (DEM)--fluid code is used to study effects of fluid injection rate on fault failure.
\item Higher fluid injection rates require higher pressures to trigger fault failure due to heterogeneous pore pressure distribution.
\item A diffusion model with a dilative sink explains failure pressure scaling with injection rate and fault length.
\end{itemize}

\section{Introduction}\label{sec:intro}

The injection of fluids into the Earth's subsurface (for energy extraction, waste disposal, or resource development) is increasingly recognised as a driver of human-induced seismicity \citep{keranen2014sharp,grigoli2017current,schultz2020hydraulic}. Fault activation by fluids is typically attributed to the reduction in fault strength caused by pore-pressure diffusion into pre-stressed faults, which lowers the effective normal stress. The relationship between pore fluid pressure $P$ and shear strength $\tau$ is commonly described using Terzaghi's effective stress law \citep{terzaghi1948soil,Hubbert1959}:
\begin{equation}
\tau = \mu(\sigma_n - P),
\label{eq:terzaghi}
\end{equation}
where $\mu$ is the friction coefficient and $\sigma_n$ is the fault-normal stress. In many applications, this framework is used with a single representative value of pore pressure, implicitly assuming that pore pressure is sufficiently uniform for a single value to control stability. However, in many practical situations, pore pressure is not uniform, making the failure criterion unclear.

The problem of heterogeneous pore pressure and its effect on failure is highlighted by recent studies \citep{Rozhko2010,aochi_self-induced_2013,French2016,passelegue2018fault,Koehn2020,Wang2020,ji2020injection,alghannam2020understanding,ji2021fluid,ji2022laboratory}, which show that the fluid pressure required to trigger fault failure can be substantially higher than predicted by Eq.~\ref{eq:terzaghi} when pore pressure is spatially heterogeneous. The experiments of \citet{passelegue2018fault}, \citet{ji2020injection} and \citet{ji2021fluid} further suggest that increasing injection rate drives a transition from relatively homogeneous to strongly heterogeneous pore-pressure distributions, so that fault reactivation is retarded by the remaining strong, low-pressure patches. Pore-pressure distribution along the fault is controlled by the competition between hydraulic diffusion and fluid injection: slow injection allows pressure to equilibrate, whereas rapid injection generates strong pressure gradients \citep{passelegue2018fault,Wang2020,ji2022high}. Rate effects also depend on the relative orientation between the injection boundary and the shear direction \citep{segall2015injection,cappa2019stabilization,cebry2021seismic}. Here we focus on the case where injection is perpendicular to shear, which maximises along-fault pressure gradients and therefore rate dependence.

Another important distinction is between bare-surface faults, such as saw-cut surfaces without gouge, and gouge-filled faults. In bare faults, injection-rate dependence is primarily controlled by fracture hydraulic connectivity, leading to sharper transitions in slip behaviour with injection rate \citep{scuderi2017frictional,passelegue2018fault,passelegue2020initial,zhao2025coupled}. In gouge-filled faults, however, failure is also affected by the granular response of the fault zone. In particular, dilation during the pre-failure phase, due to grain rearrangement,  is important in fluid-saturated faults \citep{sarma2024fault}.

 In initially over-consolidated granular layers (as in many deeply buried faults), slip is preceded by dilation \citep{reynolds1885lvii,mead1925geologic,desrues1996void,makedonska2011friction}. This dilation has two opposing effects in fluid-saturated gouge layers: First, dilation leads to more loosely packed layers that are less resistant to sliding than densely packed ones \citep{rowe1962stress,bolton1986strength,chen2016rate,chen2023emergence,sarma2025fault}, providing the weakening effect of dilation. The opposing, strengthening effect occurs during rapid dilation, which can lead to a transient pore-pressure drop and associated strengthening, or dilatant hardening \citep{frank1965dilatancy,scholz1973earthquake,segall1995dilatancy,parez2021strain,sarma2025fault}. Thus, in gouge-filled faults, injection-induced failure depends
not only on pressure diffusion but also on how the granular layer dilates as failure is approached, and the relative rates of dilation and pressurisation.

In this study, we numerically simulate fluid injection into an undrained granular layer representing a gouge-filled fault zone. The layer is pre-stressed and subjected to fluid injection at controlled rates. We observe that the pressure required for failure systematically increases with injection rate, similar to experiments on bare surfaces \citep{passelegue2018fault,ji2020injection,ji2021fluid}. In our simulations pressure is increased in discrete steps (as in the experiments of \cite{cappa2019stabilization}),  allowing equilibration between pressurisation and dilation, and thus preventing transient pressure drops. 

To explain our numerical model results, we develop an analytical model combining pore-pressure diffusion and the dilative response of the granular layer. The model predicts that failure is controlled by the least-pressurised region
of the layer, rather than by the imposed boundary pressure alone. It yields a rate-dependent failure criterion in which failure pressure increases with injection rate and fault length, and provides upper and lower bounds on the numerically observed rate dependence.

\section{Numerical Model}\label{sec:model}

We use a coupled solid--fluid code, developed in detail in previous works \citep{niebling2010mixing,goren2011mechanical,niebling2012dynamic,vass2014importance,ben2020compaction,Parez2023b}, to simulate the dynamics of grains and pore fluid in a fluid-saturated granular fault gouge subjected to increasing fluid pressure. The code combines two components: a discrete element method (DEM) for the solid grains, and a continuum pore-fluid solver for the fluid phase.

The solid phase is modelled with the discrete element method (DEM) \citep{Cundall1979}, which evolves the translational and rotational motion of the grains in two dimensions, with grains represented as disks interacting via contact forces (Section~\ref{sec:solid_phase}). DEM is widely used in geophysical and geotechnical studies ranging from earthquakes to landslides \citep{aharonov1999rigidity,aharonov2002shear,aharonov2004stick,morgan2004particle,da2005rheophysics,mair20083d,ben2010role,tordesillas2011structural,johnson2013acoustic,Parez2015,ferdowsi2015acoustically,parez2016unsteady,ferdowsi2020granular,parez2021strain,papachristos2023discrete,casas2023influence,sarma2025fault}.

The fluid phase is solved by finite differences on a coarser Eulerian grid superimposed on the grains (Section~\ref{sec:fluid_phase}). The fluid sees averaged porosity and permeability fields and is subject to the averaged deformation of the pore space within each grid cell.

\begin{figure}[ht]
\centering
\includegraphics[width=\textwidth]{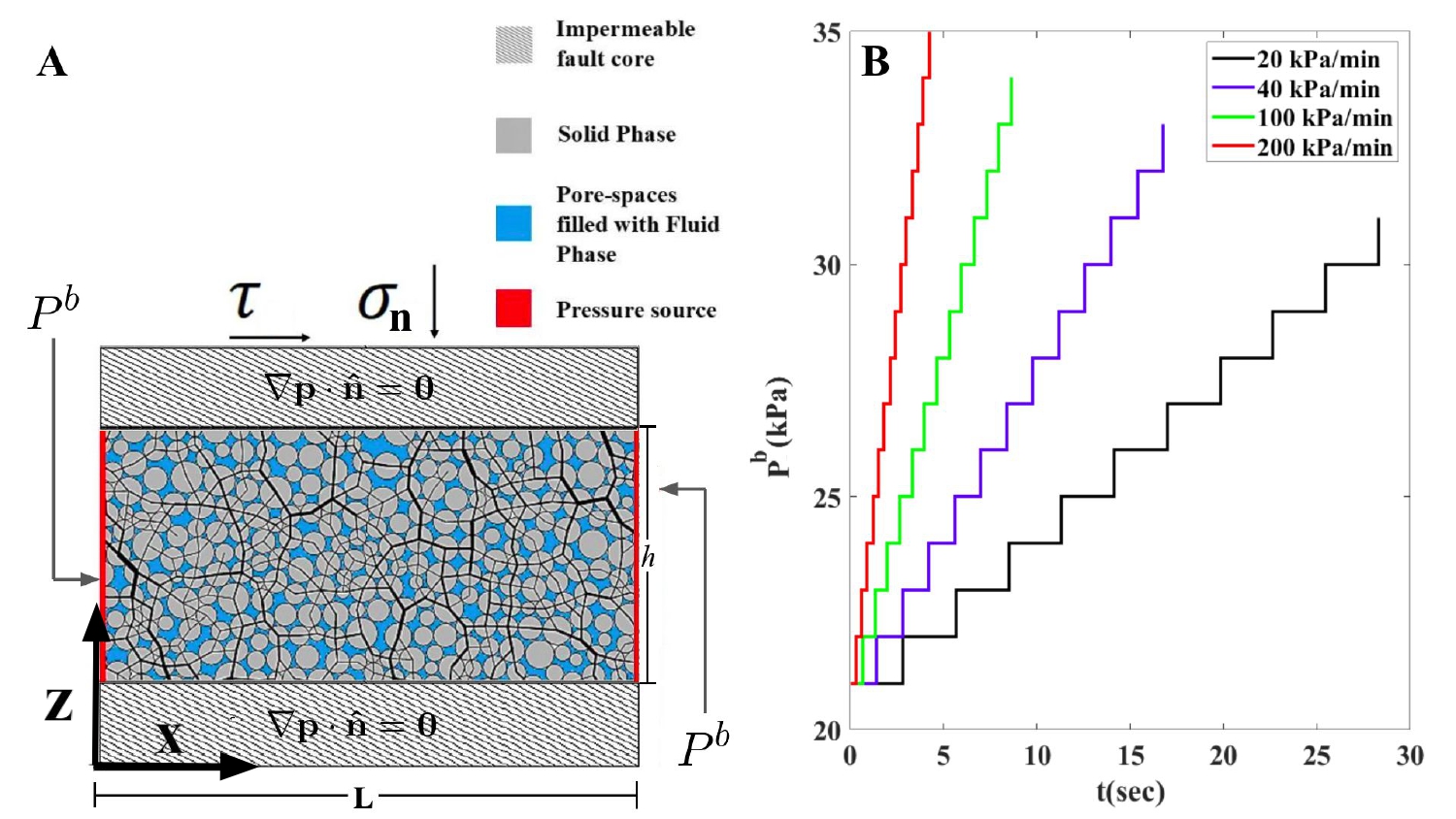}
\caption{(\textbf{A}) Numerical simulation setup: a granular layer bounded by impermeable walls, with fluid injected at prescribed pressure $P^b$ at the side boundaries (red lines). The thickness of the lines between grain centres is proportional to the intergranular contact force. (\textbf{B}) Schedule for fluid injection at four different rates between 20 and 200~kPa/min.}
\label{fig:setup_scheme}
\end{figure}

The simulation setup represents an idealised fault geometry (Fig.~\ref{fig:setup_scheme}A): a 2D granular layer of length $L$ in the $x$-direction with periodic boundary conditions, and thickness $h$ in the $z$-direction, confined between two parallel impermeable walls. The confining walls are constructed from grains glued together to emulate a rough surface. The bottom wall is static; the top wall is subject to externally applied shear stress $\tau$ and normal stress $\sigma_n$ and is free to move. The shear and normal stresses are applied first, so that the layer is pre-sheared but held below failure; fluid is then injected until failure occurs. Fluid is injected at the lateral boundaries at a controlled pressure $P^b$ (Fig.~\ref{fig:setup_scheme}A) in discrete steps, at four rates between 20 and 200~kPa/min (Fig.~\ref{fig:setup_scheme}B). Each step holds $P^b$ constant for a time $t_h$ and then increases it by $\Delta P \approx 0.01\sigma_n$; the rate is varied by changing $t_h$. The initial boundary pressure before the first step is $P^b_0 = 0.20\,\sigma_n$. This stepwise protocol follows standard experiments \citep{cappa2019stabilization} and serves as a proxy for linear pressurisation. Each simulation undergoes injection at increasing $P^b$ until the top wall fails and starts sliding under the applied constant shear stress. We record the boundary pressure at this point, $P^{b}_{\mathrm{fail}}$, and study its dependence on the rate of injection. The value of $P^{b}_{\mathrm{fail}}$ depends on pre-stress, as expected, and also on the rate of injection (Fig.~\ref{fig:pfail_overview}); to test the robustness of this rate dependence we ran simulations at multiple pre-stress levels ($\tau/\sigma_n = 0.20$--$0.25$) and layer lengths ($L/d = 72$--$384$). All simulation parameters are listed in Table~\ref{tab:params}; additional grain-scale numerical details are provided in Appendix~\ref{app:grain_details}.

\begin{table}[ht]
\centering
\caption{Parameters used in the coupled DEM--fluid simulations.}\label{tab:params}
\begin{tabular}{@{}ll@{}}
\toprule
Parameter & Value \\
\midrule
Grain density                               & $\rho_s = 2640~\mathrm{kg\,m^{-3}}$ \\
Grain Young's modulus                       & $E = 10^{8}~\mathrm{Pa}$ \\
Grain mean diameter                         & $d = 0.01~\mathrm{m}$ \\
Grain friction coefficient                  & $\mu_g = 0.5$ \\
Normal stress                               & $\sigma_n = 0.1~\mathrm{MPa}$ \\
Shear stress                                & $\tau = 0.2\text{--}0.25\,\sigma_n$ \\
Characteristic grain collision timescale    & $t_0 = 3.7\times 10^{-5}~\mathrm{s}$ \\
Hydraulic diffusivity                       & $\alpha = 0.5\,d^2/t_0$ \\
Layer length                                & $L = 72d\text{--}384d$ \\
Macroscopic friction coefficient            & $\mu = 0.21\text{--}0.27$ \\
Injection pressure                          & $P^b = 0.1\text{--}0.4\,\sigma_n$ \\
Pressure step                               & $\Delta P = 1\%\,\sigma_n$ \\
Fluid density                               & $\rho_f = 1000~\mathrm{kg\,m^{-3}}$ \\
Fluid compressibility                       & $\beta_f = 10^{-7}~\mathrm{Pa^{-1}}$ \\
Fluid dynamic viscosity                     & $\eta = 10^{-3}~\mathrm{Pa\,s}$ \\
Average permeability                        & $\kappa = 10^{-11}~\mathrm{m^{2}}$ \\
\bottomrule
\end{tabular}
\end{table}

\subsection{Solid phase}\label{sec:solid_phase}

A discrete element model with a linear elastic frictional contact law is used to solve grain dynamics \citep{Cundall1979}. Grains are represented as disks interacting via pair-wise contact forces that include elastic repulsion, velocity-dependent damping, and Coulomb friction \citep{parez2021strain,Parez2023b}. The linear and rotational momentum equations
\begin{align}
m_i\dot{\mathbf{u}}_{s,i} &= \sum_{j} \mathbf{F}_{ij} - \frac{V_i}{1-\phi}\nabla p ,\label{eq:lin}\\
I_i\dot{\boldsymbol{\omega}}_{s,i} &= \sum_{j} R_i \hat{\mathbf{n}}_{ij}\times \mathbf{F}_{ij} \label{eq:rot}
\end{align}
are time-integrated with the Verlet algorithm. Here $\dot{\mathbf{u}}_{s,i}$ and $\dot{\boldsymbol{\omega}}_{s,i}$ are the translational and rotational accelerations of grain~$i$, with mass $m_i$ and moment of inertia $I_i$. The first term on the right-hand side of Eq.~\ref{eq:lin} is the sum of contact forces $\mathbf{F}_{ij}$ with all grains $j$ in contact with grain $i$; the second is the drag force proportional to the pore-pressure gradient and grain volume $V_i$. In Eq.~\ref{eq:rot}, $R_i$ is the radius of grain $i$ and $\hat{\mathbf{n}}_{ij}$ is a unit vector from the centre of grain $i$ to that of grain $j$.

The contact force $\mathbf{F}_{ij} = \mathbf{F}^{n}_{ij} + \mathbf{F}^{t}_{ij}$ is modelled as a linear spring with velocity-dependent damping and a Coulomb friction limit:
\begin{align}
\mathbf{F}^{n}_{ij} &= -k_{n}\delta_{ij}\mathbf{n}_{ij} + \theta_{n} m_{\mathrm{eff}} \mathbf{u}^{n}_{s,ij} ,\\
\mathbf{F}^{t}_{ij} &= 
\begin{cases}
k_{t}\mathbf{t}_{ij} + \theta_{t} m_{\mathrm{eff}} \mathbf{u}^{t}_{s,ij} , & \text{if } |\mathbf{F}^{t}_{ij}| < \mu_g |\mathbf{F}^{n}_{ij}| ,\\
\mu_g |\mathbf{F}^{n}_{ij}|\,\dfrac{\mathbf{t}_{ij}}{|\mathbf{t}_{ij}|} , & \text{otherwise},
\end{cases}
\end{align}
where $k_{n,t}$ and $\theta_{n,t}$ are the stiffness and damping coefficients, $\mathbf{u}^{n}_{s,ij}$ and $\mathbf{u}^{t}_{s,ij}$ are the relative normal and tangential velocities, $m_{\mathrm{eff}}$ is the effective mass, and $\mathbf{t}_{ij}$ is the tangential displacement measured from contact formation. In the simulations, the normal damping coefficient is $\theta_n = 1.35\times 10^{5}~\mathrm{s^{-1}}$, the tangential damping coefficient is negligible ($\theta_t\approx 0$), the normal stiffness coefficient is $k_n = 10^{8}~\mathrm{Pa\,m}$, and the tangential stiffness coefficient is $k_t = 0.5 k_n = 5\times 10^{7}~\mathrm{Pa\,m}$.

\subsection{Fluid phase}\label{sec:fluid_phase}

The mechanics of the pore fluid coupled with granular motion was developed in detail in \citet{goren2010pore,goren2011mechanical} and \citet{niebling2010mixing,niebling2010sedimentation,niebling2012dynamic}. We briefly review the formulation for clarity. Mass-conservation equations for grains and fluid are
\begin{align}
\frac{\partial[(1-\phi)\rho_s]}{\partial t} + \nabla\cdot[(1-\phi)\rho_s \mathbf{u}_s] &= 0 , \label{eq:mass_solid}\\
\frac{\partial[\phi\rho_f]}{\partial t} + \nabla\cdot[\phi\rho_f \mathbf{u}_f] &= 0 , \label{eq:mass_fluid}
\end{align}
where $\rho_s$ and $\rho_f$ are the densities of the solid grains and fluid, respectively; $\mathbf{u}_s$ and $\mathbf{u}_f$ are the solid and fluid velocity fields; $\phi$ is porosity; and $t$ is time. These velocities are defined on mesoscopic volumes comprising at least a few grains.

The fluid momentum equation is approximated by Darcy's law:
\begin{equation}
\phi(\mathbf{u}_f - \mathbf{u}_s) = -\frac{\kappa}{\eta}\nabla p ,
\label{eq:darcy}
\end{equation}
where $p$ is the excess fluid pressure above hydrostatic. This formulation neglects fluid inertia and viscous drag. A linearised fluid state equation is assumed,
\begin{equation}
\rho_f = \rho_0(1+\beta_f p),
\label{eq:state}
\end{equation}
where $\beta_f$ is the adiabatic fluid compressibility and $\rho_0$ is the fluid density at the reference hydrostatic pressure. The solid phase is treated as incompressible ($\rho_s$ constant), since its compressibility is much smaller than that of the fluid.

Combining Eqs.~\ref{eq:mass_solid}--\ref{eq:state} yields a balance that also contains an advective term, $\beta_f\phi\,\mathbf{u}_s\cdot\nabla p$, representing transport of pore pressure by the moving grain skeleton. This term is negligible whenever the pore-pressure diffusion length stays larger than a grain diameter, so that pressure transport is diffusion dominated rather than advective \citep{goren2010pore,goren2011mechanical}. Neglecting it yields the governing equation for excess pore pressure \citep{goren2011mechanical,Parez2023b}
\begin{equation}
\frac{\partial p}{\partial t} - \frac{1}{\beta_f\phi\eta}\nabla\cdot[\kappa\nabla p] + \frac{1}{\beta_f\phi}\nabla\cdot \mathbf{u}_s = 0 .
\label{eq:pressure}
\end{equation}
This equation is solved on a regular grid with spacing equal to two mean grain diameters. The properties of the porous granular matrix ($\kappa$, $\phi$ and $\mathbf{u}_s$) may vary in space and are interpolated from grains onto grid points using bilinear interpolation \citep{goren2011mechanical}. Permeability and porosity are related by a Carman--Kozeny relation. Our simulation is two dimensional, so the porosity measured directly from the simulation, $\phi_{2D}$, is an areal (rather than volumetric) quantity. To estimate a 3D-equivalent permeability for the 2D simulation geometry, we follow \citet{mcnamara2000grains}, using a Carman--Kozeny-type relation for disk packings near random close packing that maps 2D porosity to 3D-equivalent permeability:
\begin{equation}
\kappa = \kappa_c \frac{(1+2\phi_{2D})^{2}}{(1-\phi_{2D})^{2}} ,
\label{eq:kozeny}
\end{equation}
where $\kappa_c$ is a prefactor that scales permeability to the desired order of magnitude.

\section{Results}\label{sec:results}

Each simulation undergoes fluid injection at a progressively increasing value of boundary pressure $P^b$ (Fig.~\ref{fig:setup_scheme}B) until the top wall fails and starts sliding under the applied constant shear stress. We record the boundary pressure at this point, $P^{b}_{\mathrm{fail}}$, and study its dependence on the rate of injection across a range of pre-stress values and layer lengths.

\begin{figure}[ht]
\centering
\includegraphics[width=\textwidth]{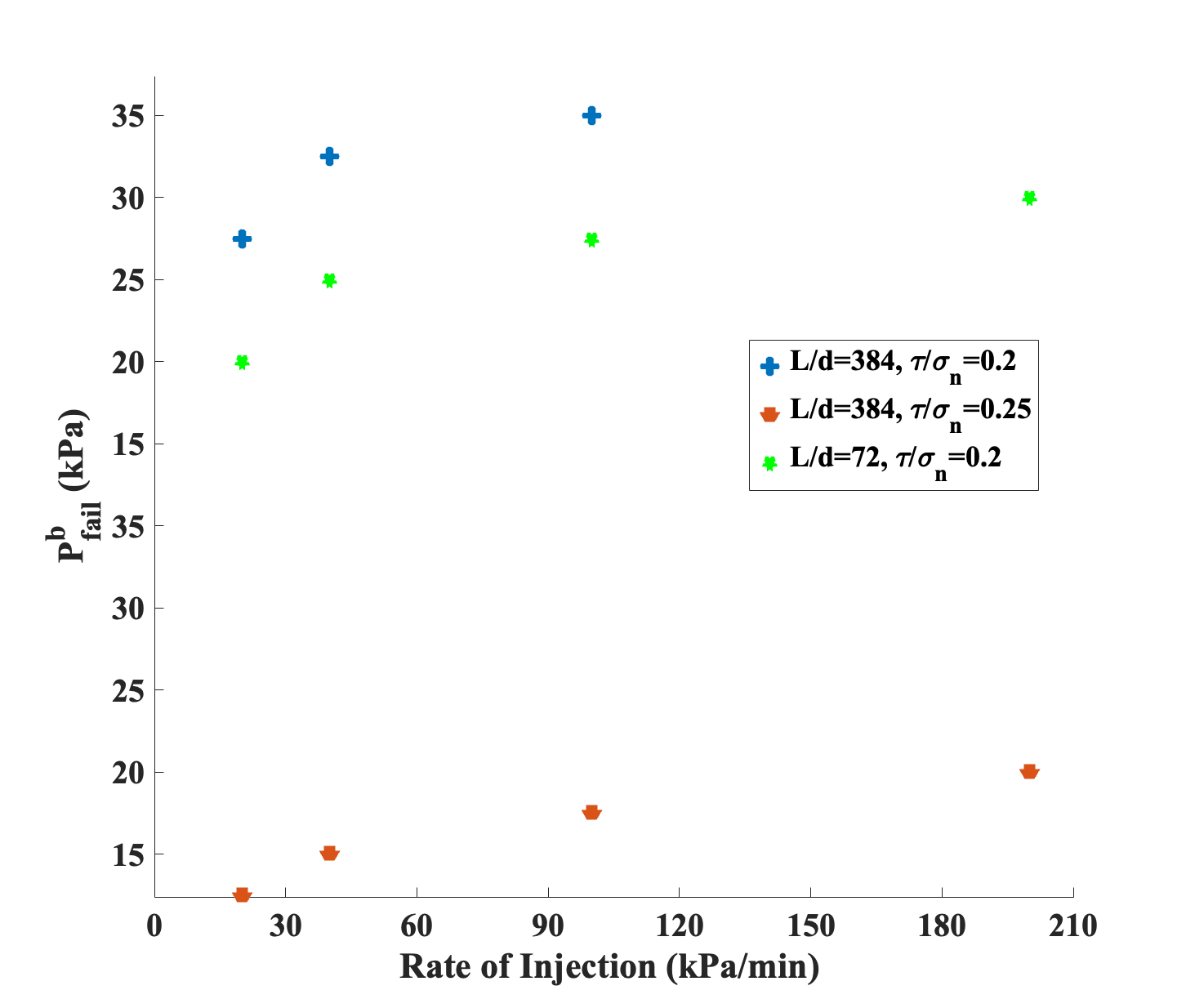}
\caption{Results of simulations showing the boundary injection pressure required to trigger failure and large-scale sliding ($P^{b}_{\mathrm{fail}}$) as a function of injection rate. Three runs are shown, with different normalised pre-stress $\tau/\sigma_n$ and normalised layer length $L/d$ (where $d$ is the mean grain diameter).}
\label{fig:pfail_overview}
\end{figure}

Across all simulations, $P^{b}_{\mathrm{fail}}$ systematically increases with injection rate (Fig.~\ref{fig:pfail_overview}). As the injection rate increases from 20 to 200~kPa/min, $P^{b}_{\mathrm{fail}}$ increases across all tested pre-stress conditions ($\tau/\sigma_n$) and normalised layer lengths ($L/d$), indicating that the effect is robust across fault geometries and stress states. $P^{b}_{\mathrm{fail}}$ also increases with system size $L$. This behaviour is consistent with experimental observations \citep{passelegue2018fault,ji2020injection,ji2021fluid} and with the theoretical prediction developed below.

\subsection{Evolution of dilation}\label{sec:dilation}

\begin{figure}[ht]
\centering
\includegraphics[width=\textwidth]{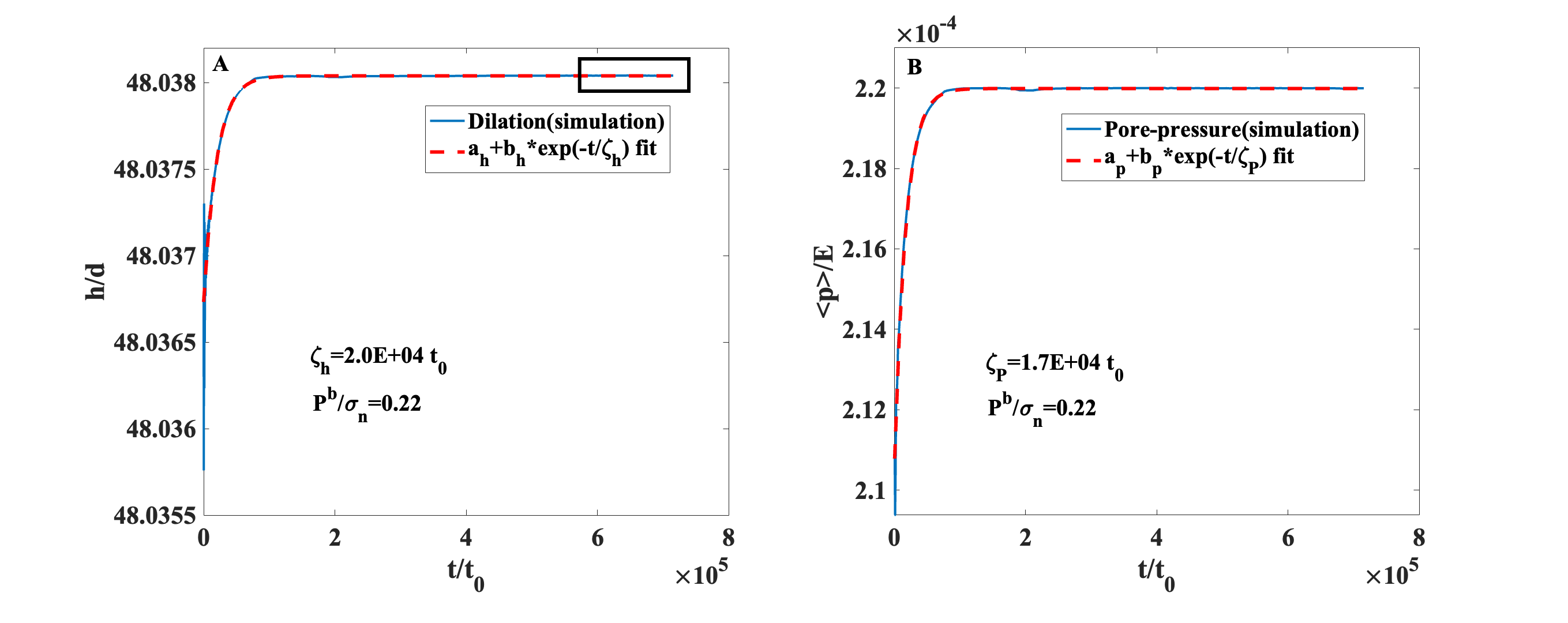}
\caption{Evolution of (\textbf{A}) layer dilation $h/d$ and (\textbf{B}) space-averaged excess pore pressure $\langle p\rangle/E$ within a single step increase of $P^b$ by $\Delta P$. Dashed lines are exponential fits $a_{h}+b_{h}\exp(-t/\zeta_{h})$ and $a_{p}+b_{p}\exp(-t/\zeta_{p})$, respectively. The black box in (A) marks the late-time window used to define the equilibrated thickness for each pressure step.}
\label{fig:equilibration}
\end{figure}

\begin{figure}[ht]
\centering
\includegraphics[width=\textwidth]{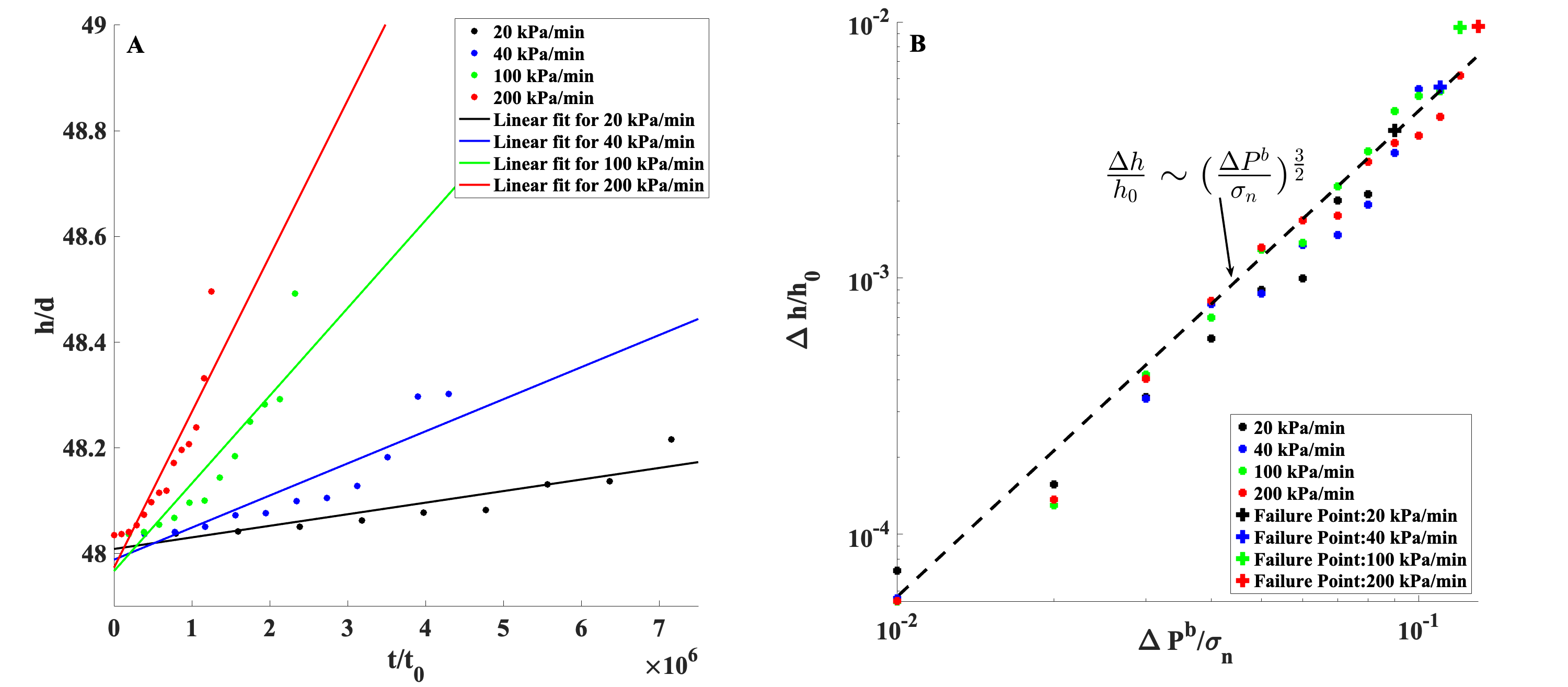}
\caption{Evolution of dilation with boundary pressurisation. (\textbf{A}) Layer thickness $h/d$ (measured at the end of each pressure step) versus normalised time $t/t_0$ for four injection rates; solid lines are linear fits. (\textbf{B}) Log--log plot of normalised dilation $\Delta h/h_0$ versus $\Delta P^{b}/\sigma_n$, showing power-law scaling with exponent $\sim 3/2$ (dashed line). Coloured stars mark the failure points at each injection rate.}
\label{fig:dilation}
\end{figure}

Following each step increase of boundary pressure, both the layer thickness $h$ and the average pore pressure $\langle p\rangle$ evolve on a similar characteristic timescale $\zeta$ \citep{sarma2025fault} towards a new quasi-equilibrium state (Fig.~\ref{fig:equilibration}). The layer thickness measured at the end of each pressure step (see box in Fig.~\ref{fig:equilibration}A) increases approximately linearly with time, in response to the approximately linear increase in $P^b$ (Fig.~\ref{fig:dilation}A). The observed increase in the rate of dilation with injection rate suggests a possible elastic response of the layer, whereby volumetric strain rate is proportional to the rate of pressurisation.

When dilation is examined as a function of pressure, however (Fig.~\ref{fig:dilation}B), all injection rates collapse onto a single curve with $\Delta h/h_0 \sim (\Delta P^b/\sigma_n)^{3/2}$, indicating that dilation grows faster than linearly with pressure. This departs from a purely elastic response and arises from progressive granular rearrangements, where  the system rearranges into a  looser packing as it approaches failure. Despite this nonlinearity, we approximate the pre-failure response as effectively linear over each pressure step and write
\begin{equation}
K\,\frac{\dot h}{h} = \dot P^{b},
\label{eq:Kdot}
\end{equation}
where $K$ is the effective bulk modulus of the grain packing. Equation~\ref{eq:Kdot} captures the general trend in Fig.~\ref{fig:dilation}A, although it does not reproduce the somewhat nonlinear pressure--dilation relation in Fig.~\ref{fig:dilation}B.

\begin{figure}[ht]
\centering
\includegraphics[width=\textwidth]{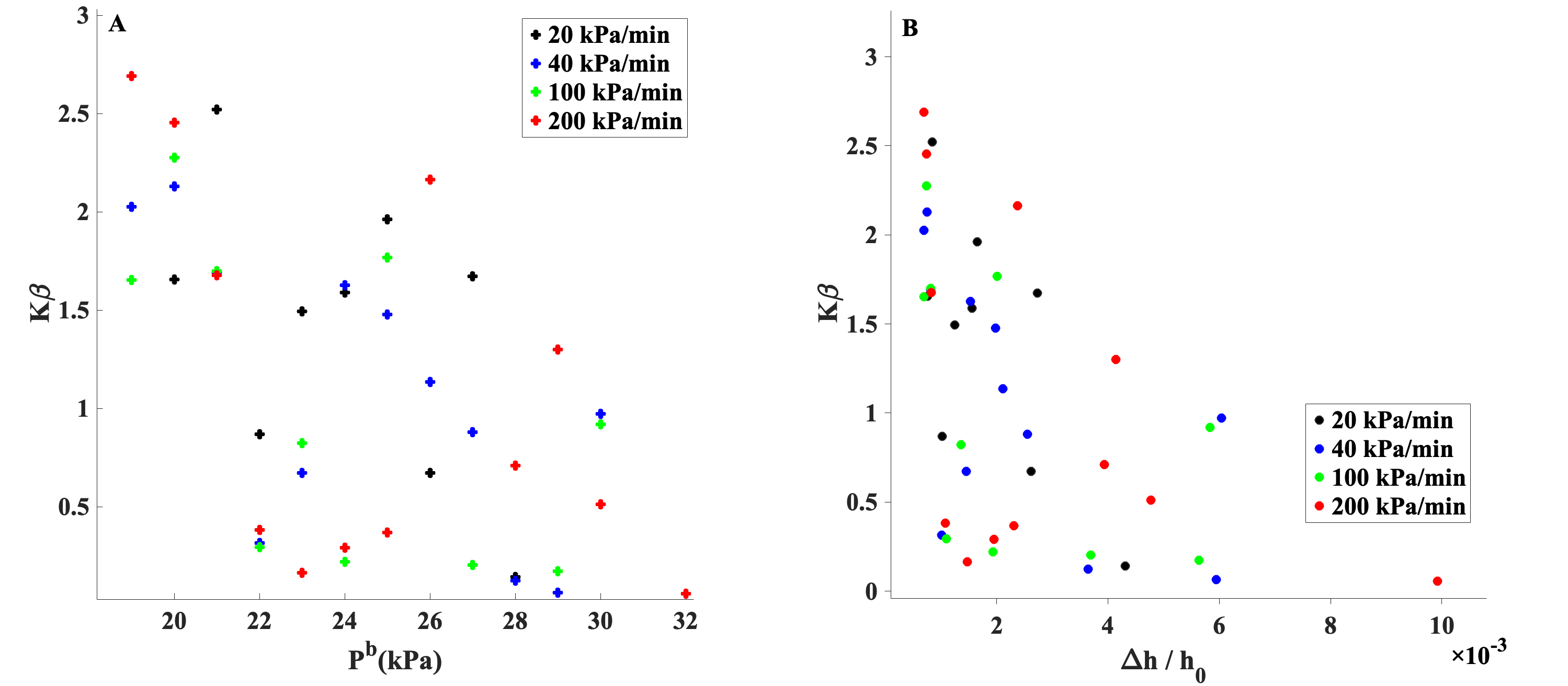}
\caption{Evolution of the effective grain-packing rigidity $K\beta$ during pressurisation. (\textbf{A}) $K\beta$ versus boundary pressure $P^{b}$ for four injection rates. (\textbf{B}) $K\beta$ versus normalised dilation $\Delta h/h_0$, showing progressive weakening of the granular packing with dilation.}
\label{fig:Kbeta}
\end{figure}

The effective modulus $K$ is estimated from the ratio of the pressure increment to the corresponding change in layer thickness, $K = \Delta P/(\Delta h/h_0)$, where $h_0$ is the initial layer thickness and $\Delta h$ and $\Delta P$ are the steady-state differences before and after each pressure step (Fig.~\ref{fig:equilibration}). As shown in Fig.~\ref{fig:Kbeta}A, $K$ is not constant but decreases with increasing $P^b$. This decrease reflects dilatancy and granular rearrangements that progressively reduce the packing rigidity as failure approaches \citep{rowe1962stress,bolton1986strength,makedonska2011friction}. The direct relationship between dilation and packing rigidity is shown in Fig.~\ref{fig:Kbeta}B, where $K\beta$ decreases with increasing $\Delta h/h_0$, confirming that dilation weakens the granular packing. Notably, the different injection rates fail at different dilation levels, with higher rates failing at larger dilation (colour-coded stars in Fig.~\ref{fig:dilation}B). For simplicity, we neglect the variation of $K$ in the analytical model below.

\subsection{Pore-pressure diffusion with a dilative sink}\label{sec:diffusion}

We now formulate a diffusion model for pore pressure within the layer during fluid injection. This formulation builds on the approximation of a linear relation between dilation rate and pressurisation rate (Eq.~\ref{eq:Kdot}), and also on a simplified representation in which dilation enters as an effective, spatially uniform sink term.

We consider pore-pressure diffusion in a 2D layer of length $L$ and thickness $h$, with impermeable boundaries at $z=0$ and $z=h$ and time-dependent pressure imposed at the lateral boundaries $x=0$ and $x=L$. The boundary pressure increases at a constant rate $\dot P^{b}$, and the solution applies to times prior to macroscopic slip.

Pore-pressure evolution is governed by a diffusion equation with a sink term arising from volumetric deformation of the granular skeleton \citep{goren2011mechanical,ben2023drainage,Parez2023b,sarma2025fault}:
\begin{equation}
\frac{\partial p}{\partial t} = \alpha\!\left(\frac{\partial^{2} p}{\partial x^{2}} + \frac{\partial^{2} p}{\partial z^{2}}\right) - \frac{1}{\beta}\!\left(\frac{\partial u_x}{\partial x} + \frac{\partial u_z}{\partial z}\right),
\label{eq:diffusion_full}
\end{equation}
where $p$ is the excess fluid pressure above hydrostatic, $\alpha = \kappa/(\eta\beta)$ is the hydraulic diffusivity, $\kappa$ is permeability, $\eta$ is fluid viscosity, $\beta = \beta_f\phi$ is the storage capacity ($\beta_f$ is fluid compressibility, $\phi$ is porosity) and $u$ is the skeleton velocity. Because the layer is laterally extensive ($L\gg h$) and bounded by impermeable walls in the vertical direction, pore pressure is nearly uniform along $z$. Integrating Eq.~\ref{eq:diffusion_full} over $z$ and defining the depth-averaged pore pressure $P(x,t) = (1/h)\int_{0}^{h} p(x,z,t)\,\mathrm{d}z$ yields
\begin{equation}
\frac{\partial P}{\partial t} = \alpha\,\frac{\partial^{2} P}{\partial x^{2}} - \frac{\dot h}{\beta h}.
\label{eq:diffusion_avg}
\end{equation}

Using Eq.~\ref{eq:Kdot} with $K$ treated as an effective constant, Eq.~\ref{eq:diffusion_avg} becomes a diffusion equation with a uniform dilative sink:
\begin{equation}
\frac{\partial P}{\partial t} = \alpha\,\frac{\partial^{2} P}{\partial x^{2}} - \frac{\dot P^{b}}{K\beta}.
\label{eq:diffusion_sink}
\end{equation}
The boundary conditions are
\begin{equation}
P(0,t) = P(L,t) = P_{0} + \dot P^{b} t,
\end{equation}
with $P_{0}$ the initial depth-averaged pore pressure and $\dot P^{b}$ the imposed constant pressurisation rate. Solving Eq.~\ref{eq:diffusion_sink} by Fourier series (see Appendix~\ref{app:diffusion_solution} for the full derivation):
\begin{equation}
P(x,t) = P_{0} + \dot P^{b} t - \frac{\dot P^{b}}{2\alpha}\!\left(1+\frac{1}{K\beta}\right) x(L-x) + \frac{4 L^{2} \dot P^{b}}{\pi^{3}\alpha}\!\left(1+\frac{1}{K\beta}\right)\!\sum_{n\,\text{odd}} \frac{\mathrm{e}^{-\alpha(\pi n/L)^{2} t}}{n^{3}}\sin\!\left(\frac{\pi n x}{L}\right).
\label{eq:P_series}
\end{equation}
For sufficiently large times ($\pi^{2}\alpha t/L^{2} > 1$) only the $n=1$ term contributes significantly:
\begin{equation}
P(x,t) \approx P_{0} + \dot P^{b} t - \frac{\dot P^{b}}{2\alpha}\!\left(1+\frac{1}{K\beta}\right) x(L-x) + \frac{4 L^{2} \dot P^{b}}{\pi^{3}\alpha}\!\left(1+\frac{1}{K\beta}\right) \mathrm{e}^{-\alpha(\pi/L)^{2} t}\sin\!\left(\frac{\pi x}{L}\right).
\label{eq:P_n1}
\end{equation}

\subsection{Predicting the failure pressure}\label{sec:prediction}

\begin{figure}[ht]
\centering
\includegraphics[width=0.75\textwidth]{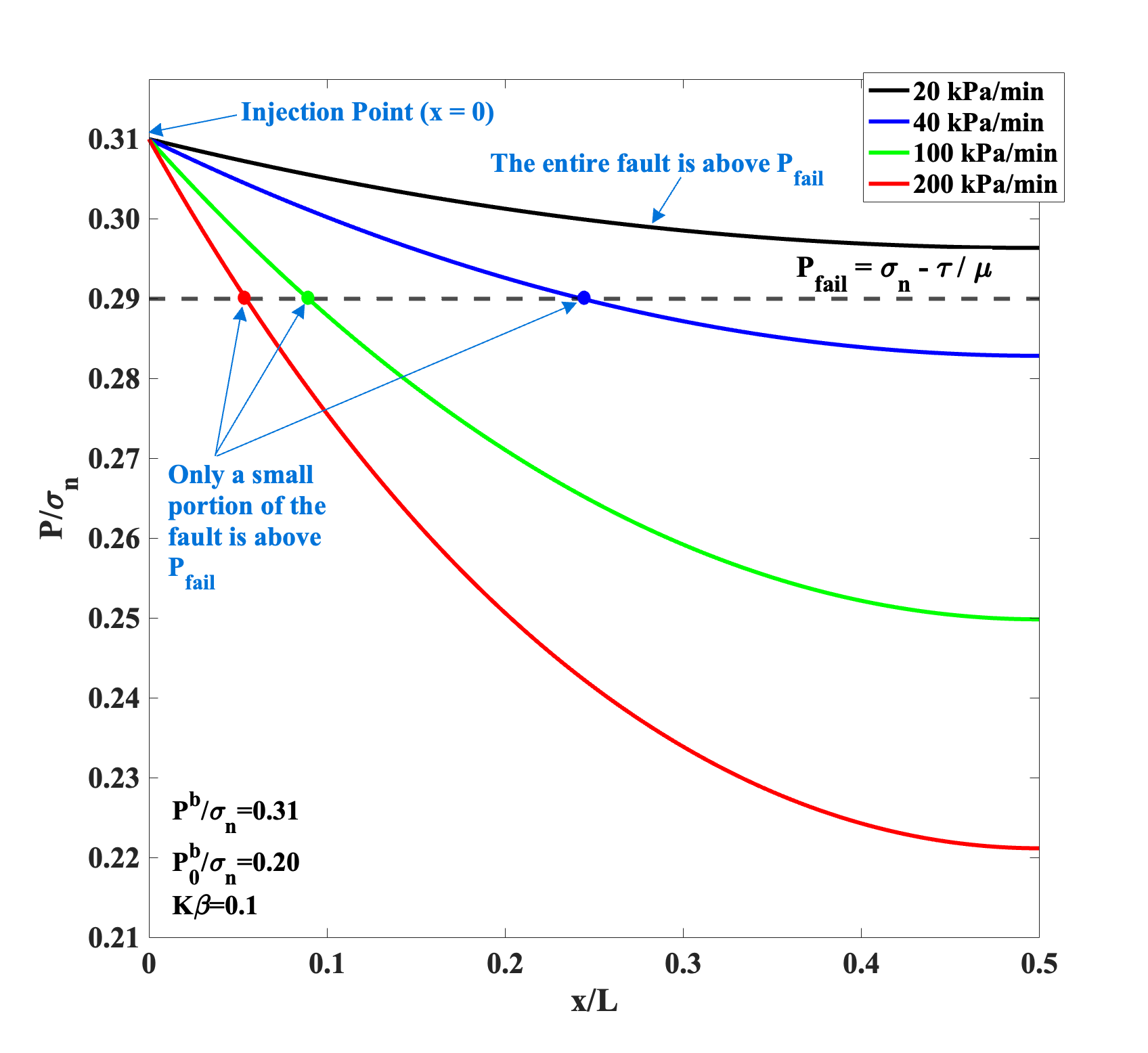}
\caption{Predicted pore-pressure profiles along the fault for the four injection rates (20, 40, 100 and 200~kPa/min), computed from the diffusion solution (Eqs.~\ref{eq:P_series} and~\ref{eq:P_n1}). Curves show the depth-averaged excess pore pressure $P/\sigma_n$ versus normalised distance $x/L$; by symmetry only the half-layer $0 \le x/L \le 0.5$ is shown. The horizontal dashed line marks the Mohr--Coulomb threshold $P_\mathrm{fail}=\sigma_n-\tau/\mu$ (coloured dots mark where each profile crosses it). Slow injection equilibrates and keeps the whole fault above $P_\mathrm{fail}$, whereas faster injection leaves only a small near-boundary region above threshold, setting the higher boundary pressure required for global failure. Parameters: $K\beta=0.1$, $\tau=0.20\,\sigma_n$, $\mu=0.28$, $L=384\,d$, $P^{b}=0.31\,\sigma_n$, $P^{b}_0=0.20\,\sigma_n$.}
\label{fig:cartoon}
\end{figure}

The diffusion solution predicts that pore pressure remains spatially heterogeneous during pressurisation, with the lowest pressure at the point farthest from the injection boundaries. Figure~\ref{fig:cartoon} shows the predicted profiles (Eqs.~\ref{eq:P_series} and~\ref{eq:P_n1}) for the four injection rates at a common boundary pressure: as the rate increases the profiles steepen and a progressively smaller fraction of the layer exceeds the Mohr--Coulomb threshold $P_\mathrm{fail}=\sigma_n-\tau/\mu$. Fault failure is therefore controlled not by the average pressure but by the least-pressurised (strongest) location within the layer. In our geometry, with injection from boundaries perpendicular to shear, the profile is symmetric and attains its minimum at $x=L/2$. Assuming global failure occurs once this last region reaches the Mohr--Coulomb criterion, we obtain a prediction for $P(x=L/2)$ at failure:
\begin{equation}
\frac{\tau}{\sigma_n - P(x=L/2)} = \mu .
\label{eq:MC}
\end{equation}
Evaluating Eq.~\ref{eq:P_n1} at $x=L/2$ then gives the boundary pressure at failure:
\begin{equation}
P^{b}_{\mathrm{fail}} = P_{0} + \dot P^{b} t_s = \sigma_n - \frac{\tau}{\mu} + \frac{\dot P^{b}}{2\alpha}\!\left(1+\frac{1}{K\beta}\right)\frac{L^{2}}{4} - \frac{4 L^{2} \dot P^{b}}{\pi^{3}\alpha}\!\left(1+\frac{1}{K\beta}\right)\mathrm{e}^{-\alpha(\pi/L)^{2} t_s},
\label{eq:pfail_full}
\end{equation}
where $t_s$ is the time to failure since the start of injection. For high hydraulic diffusivity, $\alpha(\pi/L)^{2} t_s \gg 1$, the exponential term can be neglected and
\begin{equation}
P^{b}_{\mathrm{fail}} \approx \sigma_n - \frac{\tau}{\mu} + \frac{L^{2} \dot P^{b}}{8\alpha}\!\left(1+\frac{1}{K\beta}\right).
\label{eq:pfail}
\end{equation}
Equation~\ref{eq:pfail}, derived for constant $K$, is used to bound the simulation results, which show variation of $K$ across three orders of magnitude (see Fig.~\ref{fig:Kbeta} for the origin of this variation, discussed in Section~\ref{sec:dilation}).

\begin{figure}[ht]
\centering
\includegraphics[width=0.85\textwidth]{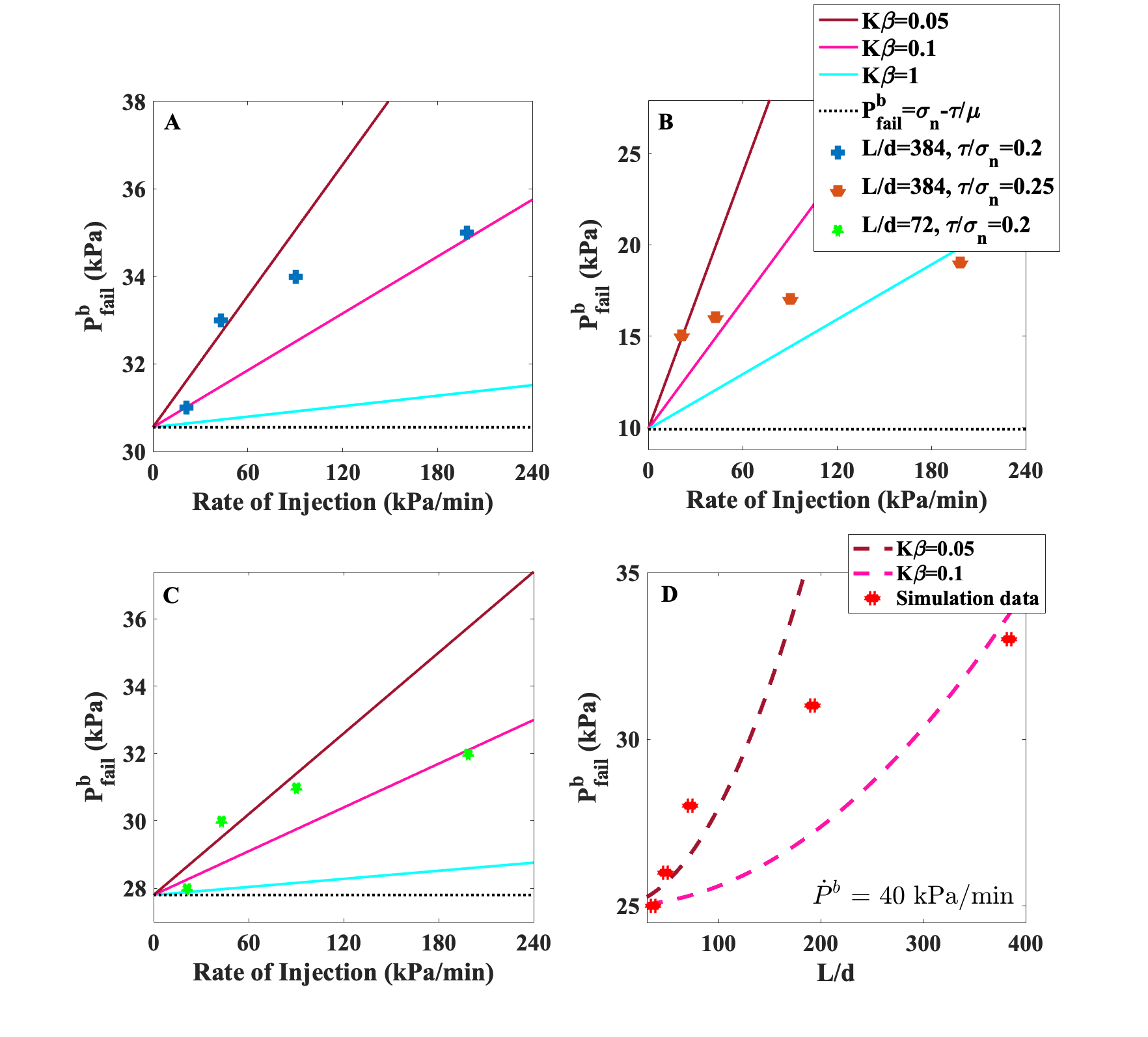}
\caption{Comparison between simulations (symbols) and the analytical prediction (Eq.~\ref{eq:pfail}, solid lines). (\textbf{A--C}) Failure pressure $P^{b}_{\mathrm{fail}}$ versus injection rate for three representative cases of normalised layer length $L/d$ and pre-stress $\tau/\sigma_n$, using effective values of $K\beta$ spanning three orders of magnitude (see Fig.~\ref{fig:Kbeta}). The horizontal dotted line indicates the uniform-pressure critical value $P^{b}_{\mathrm{fail}} = \sigma_n - \tau/\mu$. (\textbf{D}) Failure pressure versus normalised fault length $L/d$ for a fixed injection rate $\dot P^{b} = 40$~kPa/min. Dashed lines show analytical predictions for two values of $K\beta$.}
\label{fig:theory}
\end{figure}

\newpage
Figure~\ref{fig:theory}A--C compares Eq.~\ref{eq:pfail} (solid lines) with simulation results (symbols) for effective values of $K\beta$ ranging from 0.05 to 1. The theoretical predictions provide robust bounds on the relationship between injection rate and failure pressure for three representative layer lengths and pre-stress levels, with lower $K\beta$ values bounding the observed failure pressures more tightly. Equation~\ref{eq:pfail} also predicts a strong $L^{2}$ dependence of $P^{b}_{\mathrm{fail}}$ on fault length. Although the simulations do not show such a strong dependence, failure pressure does increase with $L$, and the analytical model successfully bounds the numerical observations (Fig.~\ref{fig:theory}D).

\section{Discussion}\label{sec:discussion}

This work shows that rate-dependent fault failure during fluid injection arises from diffusion-controlled pore-pressure heterogeneity (Fig.~\ref{fig:cartoon}). When injection is slow, pore pressure equilibrates along the fault, leading to nearly uniform weakening and failure at low boundary pressures. Rapid injection instead produces strong spatial pressure gradients: regions near the injection boundary weaken while more distant regions remain strong, so failure is controlled by the least-pressurised location, requiring higher boundary pressures for the entire fault to fail. Our analytical model captures this by combining pore-pressure diffusion with a dilative sink term, yielding a modified failure condition (Eq.~\ref{eq:pfail}) in which failure pressure increases with injection rate and scales with the square of fault length. This analysis depends on  fault vs. injection orientation: it applies when injection is perpendicular to shear, so that pressure must diffuse along the fault and failure is controlled by the least-pressurised central region. In contrast, injection parallel to shear creates a continuous shear band adjacent to the injection line as soon as pressure reaches the Mohr--Coulomb threshold, making failure largely independent of diffusion rates.

In addition to pore-pressure heterogeneity that arises from gradients in pore-pressure diffusion, dilation also influences layer strength and pore pressure: In dense fault gouge, sliding is preceded by dilation \citep{reynolds1885lvii,mead1925geologic,marone1990frictional,makedonska2011friction}. This dilation influences the path to failure in two opposite ways: the first is a strengthening effect---rapid dilation can outpace pore-pressure diffusion, producing a transient pore-pressure drop and associated strengthening (dilatant hardening) \citep{frank1965dilatancy,scholz1973earthquake,rice1975stability,segall1995dilatancy}. When dilation occurs on timescales comparable to or slower than pressure homogenisation, pore pressure remains near equilibrium and this transient strengthening is negligible. The second effect of dilation is that of weakening, and it is independent of pore-pressure. Dilation via irreversible grain rearrangements progressively reduces packing density as the system approaches failure. Since there is a direct correlation between packing density and shear strength, the layer becomes progressively weaker as it dilates on its route to failure \citep{rowe1962stress,bolton1986strength,chen2016rate,chen2023emergence,sarma2025fault}.

Our simulations are controlled by this latter regime: pressure is increased in discrete steps, and both pore pressure and layer thickness equilibrate to a quasi-steady state after each step (Fig.~\ref{fig:equilibration}), preventing transient pressure drops and dilatant hardening. As a result, dilation contributes primarily through progressive weakening. In our model (Eq.~\ref{eq:pfail}), this is captured by the dilative sink term (Eq.~\ref{eq:diffusion_sink}). However, dilation also modifies material properties: as porosity increases, both $K$ and $\mu$ decrease. Incorporating porosity-dependent friction and moduli leads to 

\begin{equation}
\label{eq:pfail_phi}
 P_{fail}^{b} \approx \sigma_n - \frac{\tau}{\mu(\Delta h/h)} +  \frac{L^2 \dot{P}^b}{8 \alpha} \left(1 + \frac{1}{K(\Delta h/h) \beta} \right) \,.
\end{equation}

As dilation progresses and $K\beta$ decreases (Fig.~\ref{fig:Kbeta}B), the last term \emph{increases}. In contrast, since $\mu$ also decreases with porosity \citep{makedonska2011friction,chen2016rate}, the growing $\tau/\mu$ term \emph{reduces} the predicted failure pressure. The simulation results (Fig.~\ref{fig:theory}A--C) show sub-linear behaviour, indicating that dilation-induced friction weakening dominates, but a full treatment of porosity-dependent $\mu$ and $K$ is beyond this work's scope. We further note that $K$ depends on system size and dimensionality \citep{kuhn2009specimen,huang2014effect}, so our 2D geometry likely overestimates stiffness relative to three-dimensional fault zones.

Our framework also predicts a strong dependence of failure pressure on fault length ($P_{fail}^{b} \sim L^2$, Eq.~\ref{eq:pfail}). Although simulations do not reproduce this strong scaling, failure pressure does increase with $L$ and our model bounds the observations (Fig.~\ref{fig:theory}D). The discrepancy likely reflects assumptions of constant $K$, $\mu$, and $\alpha$. Nevertheless, these length effects imply that failure criteria cannot be extrapolated from laboratory to field scales without accounting for diffusion-controlled scaling. In natural faults, the relevant length scale may be the rupture nucleation length rather than the entire fault length \citep{dieterich1992earthquake,rubin2005earthquake,scholz2019mechanics}, which emerges as a critical control parameter on injection-induced seismicity.

\subsection{Implications for field scale fluid injection}\label{sec:field_implications}

From an operational standpoint, Eq.~\ref{eq:pfail} suggests that operators can potentially reduce fault reactivation risk by controlling injection protocols. Although faster injection raises $P^{b}_{\mathrm{fail}}$, failure occurs sooner (from Eq.~\ref{eq:pfail_full}, $t_s$ decreases with increasing $\dot P^{b}$). Because $P^{b}_{\mathrm{fail}}$ increases with injection rate, however, higher rates may also permit a larger cumulative pressure rise before reactivation; whether this translates into a larger injected fluid volume depends on injection geometry and the well/aquifer storage characteristics, which lie outside our 2D fault-scale model. This rate--pressure trade-off may nonetheless be advantageous when combined with cyclic injection--withdrawal protocols \citep{zang2019cyclic,kwiatek2019controlling}. In the field, faults are oriented at random angles to the injection pressure front \citep{ellsworth2013injection,segall2015injection,chang2020operational}, which will generally reduce the rate effect from its maximum in our perpendicular-injection setting. Our analysis also rests on idealisations that may not hold at field scales: constant hydraulic properties, whereas real faults exhibit substantial heterogeneity \citep{bense2013fault,vidal2015permeable,cappa2019stabilization,yang2021effect,curzi2023constraints,bjornaraa2023characterizing}; simplified 2D geometry versus segmented, non-planar natural faults \citep{walsh2003formation,ando2004dynamic}; and neglect of nonlinear feedbacks such as permeability evolution from mineral precipitation or clay swelling \citep{little2009late,marin2023fluid,zhenhao2023anomalous,petrie2024development}. Thermal effects could further modify the magnitude and timescale of dilatancy at field scales \citep{segall1995dilatancy,parez2021strain}. Field application of this rate-dependent framework should therefore be supplemented with site-specific geophysical monitoring and more sophisticated models accounting for heterogeneity and complexity of natural fault systems \citep{jia2018did,schultz2020hydraulic}.

Fault length is itself a key operational variable. Because $P^{b}_{\mathrm{fail}}$ scales as $L^{2}$ in our analytical model (Eq.~\ref{eq:pfail}), longer faults are predicted to be substantially harder to reactivate at a given injection rate, or, equivalently, to tolerate higher rates without failure. Shorter faults are more readily reactivated and behave closer to the uniform-pressure limit. Although our simulations show a weaker-than-$L^{2}$ scaling (Fig.~\ref{fig:theory}D), possibly reflecting the simplifying assumptions of constant $K$, $\mu$ and $\alpha$, the qualitative ordering is robust: long-fault systems should offer a wider safe operational window for fluid injection than short, critically stressed faults. At field conditions the relevant length scale may not be the entire fault length but the rupture nucleation length \citep{dieterich1992earthquake,rubin2005earthquake,scholz2019mechanics}; substituting the nucleation length for $L$ in Eq.~\ref{eq:pfail} provides a more conservative bound for natural systems.

\section{Conclusions}\label{sec:conclusions}

In summary, the relationship between fluid-pressurisation rate and failure pressure is similar for granular fault gouge and for saw-cut laboratory samples: higher injection rates produce strong pressure gradients, so that the effective normal stress is not uniformly reduced, failure is delayed by remaining strong zones, and higher boundary pressures are required for reactivation. The proposed diffusion model with a dilative sink (Eq.~\ref{eq:pfail}) predicts a rate-dependent effective-stress criterion in which the correction term scales linearly with pressurisation rate and quadratically with fault length, while varying inversely with hydraulic diffusivity and depending on the product $K\beta$. The analytical predictions successfully bound our simulations across three orders of magnitude of grain-packing rigidity $K\beta$. The simulations exhibit a sub-linear dependence of $P^{b}_{\mathrm{fail}}$ on injection rate, weaker than the linear analytical prediction, which we attribute to progressive weakening of the friction coefficient as the granular packing dilates toward failure. Finally, faster injection rates raise $P^{b}_{\mathrm{fail}}$ and may therefore allow operators to inject a larger cumulative fluid volume before reactivation, depending on injection geometry and storage, a rate--pressure trade-off with practical implications for subsurface operations where maximising injected volume while managing seismic risk is a key objective.

\section*{Declarations}

\noindent\textbf{Funding.} PS and EA acknowledge the support of the Bi-national Israel--US Industrial Development Fund of the US--Israel Energy Center for Fossil Energy and ISF grant 1261/2023. RT acknowledges the support of the University of Oslo, the Njord Centre, the CNRS IRP D-FFRACT and the Research Council of Norway through the PoreLab Center of Excellence (project number 262644). SP acknowledges the support of the Johannes Amos Comenius Programme (P~JAC), project No.~CZ.02.01.01/00/22\_008/0004605, Natural and Anthropogenic Georisks.

\noindent\textbf{Conflicts of interest.} The authors declare no competing interests.

\noindent\textbf{Availability of data and materials.} The simulation software is available at \citet{sarma_2024_13765085} under a CC BY 4.0 license. Datasets for the simulation results are available at \citet{sarma_2026_18922669}. Raw numerical data were post-processed in MATLAB R2023b; post-processing scripts are distributed with the simulation codes.

\noindent\textbf{Author contributions.} All authors contributed to the conception and design of the study. PS carried out the numerical simulations. The analytical derivation was initially drafted by RT, and was refined collaboratively with SP, EA, and PS. PS wrote the first draft of the manuscript; all authors contributed to revisions and approved the final version.

\appendix

\section{Solution to pore-pressure diffusion with a dilative sink}\label{app:diffusion_solution}

Equation~\ref{eq:diffusion_avg} of the main text is the pressure diffusion formulation with a sink term that is approximately constant (Fig.~\ref{fig:dilation}A) and is reproduced here for completeness:
\begin{equation}
\frac{\partial P}{\partial t} = \alpha\frac{\partial^{2} P}{\partial x^{2}} - \frac{\dot h}{\beta h} .
\label{eq:app_main}
\end{equation}
We seek the solution satisfying the boundary conditions
\begin{equation}
P(0,t) = P(L,t) = P_{0} + \dot P^{b} t ,
\end{equation}
with $\dot P^{b}$ the imposed pressurisation rate and the initial condition $P(x,0) = P_{0}$. Decompose $P(x,t) = P_{0} + \dot P^{b} t + v(x,t)$ into a term that satisfies the boundary condition and an unknown function $v$ satisfying
\begin{equation}
\frac{\partial v}{\partial t} = \alpha\frac{\partial^{2} v}{\partial x^{2}} - \frac{\dot h}{\beta h} - \dot P^{b} ,
\label{eq:app_v}
\end{equation}
with homogeneous boundary conditions
\begin{equation}
v(0,t) = v(L,t) = 0
\end{equation}
and initial condition $v(x,0) = 0$.

We solve Eq.~\ref{eq:app_v} by Fourier series expansion for $v$ and for the inhomogeneous term:
\begin{align}
v(x,t) &= \sum_{n=1}^{\infty} a_n(t) \sin\!\left(\frac{\pi n x}{L}\right) ,\\
\dot P^{b} + \frac{\dot h}{\beta h} &= \sum_{n=1}^{\infty} b_n \sin\!\left(\frac{\pi n x}{L}\right) ,
\end{align}
in which only sine terms contribute because of the boundary conditions. The Fourier coefficient $b_n$ for the spatially uniform inhomogeneous term is
\begin{equation}
b_n = 2\!\left(\dot P^{b}+\frac{\dot h}{\beta h}\right)\frac{1-(-1)^{n}}{\pi n} ,
\end{equation}
and is therefore non-zero only for odd $n$. Inserting the Fourier series into Eq.~\ref{eq:app_v} leads to the decoupled ODEs
\begin{align}
\dot a_n + \alpha\!\left(\frac{\pi n}{L}\right)^{2} a_n + \frac{4}{\pi n}\!\left(\dot P^{b}+\frac{\dot h}{\beta h}\right) &= 0 \quad [n\text{ odd}] ,\\
\dot a_n + \alpha\!\left(\frac{\pi n}{L}\right)^{2} a_n &= 0 \quad [n\text{ even}] ,
\end{align}
with solutions satisfying $a_n(0) = 0$:
\begin{align}
a_n(t) &= -\frac{4 L^{2}}{(\pi n)^{3}\alpha}\!\left(\dot P^{b}+\frac{\dot h}{\beta h}\right)\!\left[1 - \mathrm{e}^{-\alpha(\pi n/L)^{2} t}\right] \quad [n\text{ odd}] ,\\
a_n(t) &= 0 \quad [n\text{ even}] .
\end{align}
Combining these gives the depth-averaged pore-pressure evolution
\begin{equation}
P(x,t) = P_{0} + \dot P^{b} t - \frac{4 L^{2}}{\pi^{3}\alpha}\!\left(\dot P^{b}+\frac{\dot h}{\beta h}\right)\sum_{n\,\text{odd}} \frac{1-\mathrm{e}^{-\alpha(\pi n/L)^{2} t}}{n^{3}}\sin\!\left(\frac{\pi n x}{L}\right) .
\label{eq:app_P_series}
\end{equation}
Using the Fourier series expansion of $x(L-x)$, Eq.~\ref{eq:app_P_series} can be recast as
\begin{equation}
P(x,t) = P_{0} + \dot P^{b} t - \frac{1}{2\alpha}\!\left(\dot P^{b}+\frac{\dot h}{\beta h}\right) x(L-x) + \frac{4 L^{2}}{\pi^{3}\alpha}\!\left(\dot P^{b}+\frac{\dot h}{\beta h}\right)\!\sum_{n\,\text{odd}}\frac{\mathrm{e}^{-\alpha(\pi n/L)^{2} t}}{n^{3}}\sin\!\left(\frac{\pi n x}{L}\right) ,
\end{equation}
which yields Eq.~\ref{eq:P_series} of the main text when combined with Eq.~\ref{eq:Kdot}.

\section{Grain-scale numerical details}\label{app:grain_details}

Grain diameters follow a Gaussian distribution with mean $d = 0.01$~m and a maximum polydispersity of $\pm 20\%$. Grain size is set larger, and Young's modulus smaller, than in natural gouge \citep{Billi2005} to expedite numerical simulations. DEM requires resolving grain collisions occurring on the timescale $t_0 = d\sqrt{\pi\rho_s/(6E)}$ for mean grain diameter $d$, grain Young's modulus $E$, and grain mass density $\rho_s$. Granular rearrangements leading to layer failure occur on much longer timescales and are insensitive to the details of grain-collision dynamics. For numerical efficiency, we therefore artificially increase $t_0$, effectively simulating a medium with a slower speed of sound.

\bibliographystyle{plainnat}
\bibliography{bibliography}

\end{document}